\documentclass[12pt,tightenlines]{revtex4-1}   
\usepackage{graphicx}
\usepackage{hyperref}
\hypersetup{
  pdfauthor={Florian Sedlmeir},
  pdftitle={Experimental characterization of an uniaxial angle cut whispering gallery mode resonator},
  pdfsubject={ Resonators;  Optical Resonators;  Crystal Optics;  Birefringence},
  urlcolor=blue,
}

\newcommand{\mgf} {MgF$_2$ }
\newcommand{\mgfns} {MgF$_2$}

\newcommand{\LTns} {LiTaO$_3$}
\newcommand{\LNns} {LiNbO$_3$}

\newcommand{\unit}[1]{\ensuremath{\, \mathrm{#1}}}

\begin{document}

\title{Experimental characterization of an uniaxial angle cut whispering gallery mode resonator}
\author{Florian Sedlmeir,$^{1,2,3}$ Martin Hauer,$^{1}$ Josef U. F\"urst,$^{1,2}$  \\Gerd Leuchs,$^{1,2}$ and Harald G. L. Schwefel$^{1,2}$}
\address{$^1$Institute for Optics, Information and Photonics, \\University of Erlangen-N\"urnberg, 91058 Erlangen, Germany\\
$^2$Max Planck Institute for the Science of Light,
91058 Erlangen, Germany\\
$^3$SAOT, School in Advanced Optical Technologies, \\Paul-Gordan-Str.\ 6,
91052 Erlangen, Germany}
\email{Florian.Sedlmeir@mpl.mpg.de}
\author{\today}

\begin{abstract}
The usual configuration of uniaxial whispering gallery mode resonators is a disk shaped geometry where the optic axis points along the symmetry axis, a so called z-cut resonator. Recently x-cut resonators, where the optic axis lies in the equatorial plane, became of interest as they enable extremely broadband second harmonic generation. In this paper we report on the properties of a more generalized system, the so called angle-cut resonator, where the optic axis exhibits an arbitrary angle against the symmetry axis. We show experimentally that the modal structure and quality factors are similar to common resonators but that the polarization properties differ quite significantly: due to the asymmetry the polarization depends on the equatorial position and is, in general, elliptical.
\end{abstract}
\maketitle


\section{Introduction}

In their simplest and most usual form, whispering gallery mode (WGM) resonators are made from isotropic materials. Isotropy is inherent to amorphous WGM resonators such as droplets \cite{benner_observation_1980}, molten silica spheres \cite{braginsky_quality-factor_1989}, or toroids \cite{armani_ultra-high-q_2003}. Moreover, all crystals with cubic symmetry like\ CaF$_2$ are isotropic. They have found applications in ultra high $Q$ WGM resonators \cite{savchenkov_optical_2007} which are used, for example, as filtered feedback for ultra narrow lasers \cite{liang_whispering-gallery-mode-resonator-based_2010,sprenger_caf2_2010,collodo_sub-khz_2012}. Most of the crystals fall, however, in one of the four other symmetry classes, where the refractive index varies, depending on polarization and direction of the light propagation. 
Among crystals of these four symmetry classes, the group of uniaxial crystals show only one direction of light propagation, i.e. one \textit{optic axis}, where the refractive index is independent of the polarization. The propagation of light in the plane perpendicular to the optic axis is then governed by the ordinary index of refraction $n_o$ for perpendicular polarization, while the extraordinary index of refraction $n_e$ is governed by the polarization parallel to the optic axis.

Predominant uniaxial materials that have been used for WGM resonators are \LNns, \LTns, crystal quartz and magnesium fluoride (\mgfns). The former three were used as electro-optic modulators \cite{ilchenko_whispering-gallery-mode_2003,savchenkov_tunable_2009,ilchenko_crystal_2008}, for second harmonic generation \cite{ilchenko_nonlinear_2004, furst_naturally_2010}, parametric down-conversion \cite{savchenkov_parametric_2007, furst_low-threshold_2010,beckmann_highly_2011}, and recently for non-classical light generation \cite{furst_quantum_2011,fortsch_versatile_2013}. The latter one was used for opto-mechanical studies\cite{hofer_cavity_2010}, narrow line-width frequency standards \cite{alnis_thermal-noise-limited_2011}, and ultra-precise temperature measurement and stabilization \cite{strekalov_temperature_2011}. Except in \cite{ilchenko_crystal_2008} where an x-cut quartz resonator was used, WGM resonators are usually cut such that its axis of rotation coincides with the optic axis, the so called z-cut configuration. Thus, for high $Q$ resonances that are closely confined at the equator, the modes would be either polarized parallel (TE) to the optic axis, or perpendicular (TM), seeing either the extraordinary or ordinary index. There has recently been a report that such a TM polarized WGM provides a significant amount of field parallel to the propagation of the mode even in isotropic WGM resonators \cite{junge_strong_2013}.

It was shown recently that the special case of an x-cut resonator can be utilized as a very broadbanded nonlinear device for second harmonic generation \cite{lin_broadband_2013}. In this case, TE and TM modes still exist and correspond to ordinary and extraordinary eigenpolarization of the crystal. However, contrary to z-cut resonators, the TM mode experiences a continuously varying group velocity depending on the angular position on the resonators rim: it oscillates between $n_o$ and $n_e$. Therefore, due to dispersion of pump and signal wavelengths there are four phase-matched points for TE and TM light over a very broad wavelength regime. Though the phase-matching takes place only in a very small region of the modes, reasonable conversion efficiencies can be achieved due to the high quality factors of WGM resonators. So far second harmonic generation from pump wavelengths of $637 \unit{nm}$ to $1557 \unit{nm}$ with conversion efficiencies up to $4.6 \unit{\%/mW}$ was demonstrated in an x-cut BBO WGM resonator \cite{lin_broadband_2013}.

To enhance this conversion efficiency one can go for a tradeoff between bandwidth and length of the phase-matched regions: fabricating the resonator in a way that the optic axis has some angle between $0^\circ$ and $90^\circ$ against the symmetry axis will decrease the modulation amplitude of the refractive index for the extraordinary WGM mode and therefore decrease the tunability of the system. At the same time, however, the length of the phase-matched regions will increase, giving rise to improved conversion efficiency. 
The question arises, whether such an asymmetric system actually shows modes and if yes, what kind of polarization behavior they will have.

We thus report on an experimental investigation of such a WGM resonator cut out of the uniaxial crystal \mgfns, so that its symmetry axis makes a nonzero angle with the optic axis. Contrary to \cite{lin_high-q_2012}, where BBO was used, we have a resonator made of positive crystal showing only weak birefringence. Furthermore we provide a throughout investigation of the polarization states of the modes by Stokes analysis in transmission. Indeed it turns out that the resonator shows different types of high-Q modes which greatly differ from modes in common z-cut resonators in terms of their polarization behavior.

\section{Theoretical Considerations}
Modeling such a system theoretically is challenging. As the system lags rotational symmetry the fully vectorial Maxwells equation needs to be solved (numerically) in three dimensions. For a realistic millimeter sized resonator in the optical domain this is currently not feasible. Initial simulations for much smaller resonators show interesting results but extrapolation to the larger resonators studied here is not straight forward. In the short wavelength limit, ray-optical approaches could provide some insight. Reflections at boundaries with arbitrary oriented optic axis lead to a number of interesting effects: in general, the incoming ray splits into two rays, having ordinary and extraordinary polarization as well as different propagation directions. The energy distribution between these two rays is described by generalized Fresnel coefficients \cite{sluijter_general_2008}. For close to glancing incidence the so called \textit{inhibited reflection} \cite{simon_inhibited_1989} can occur. In a positive crystal, where $n_e > n_o$, an incident extraordinary polarized ray will no longer couple to the ordinary polarization and only one (extraordinary polarized) reflected ray will appear. As this is not the case for incident ordinary polarized light it was predicted \cite{grudinin_polarization_2013} that an ordinary polarized WGM should decay due to multiple reflections at the resonators rim. Therefore such a WGM should either not exist or having a reduced $Q$ factor compared to the extraordinary WGM. A theory that takes all this into consideration, is beyond the scope of this publication.

\section{Experimental Setup}

\begin{figure}[t]
\centering\includegraphics[height=6cm]{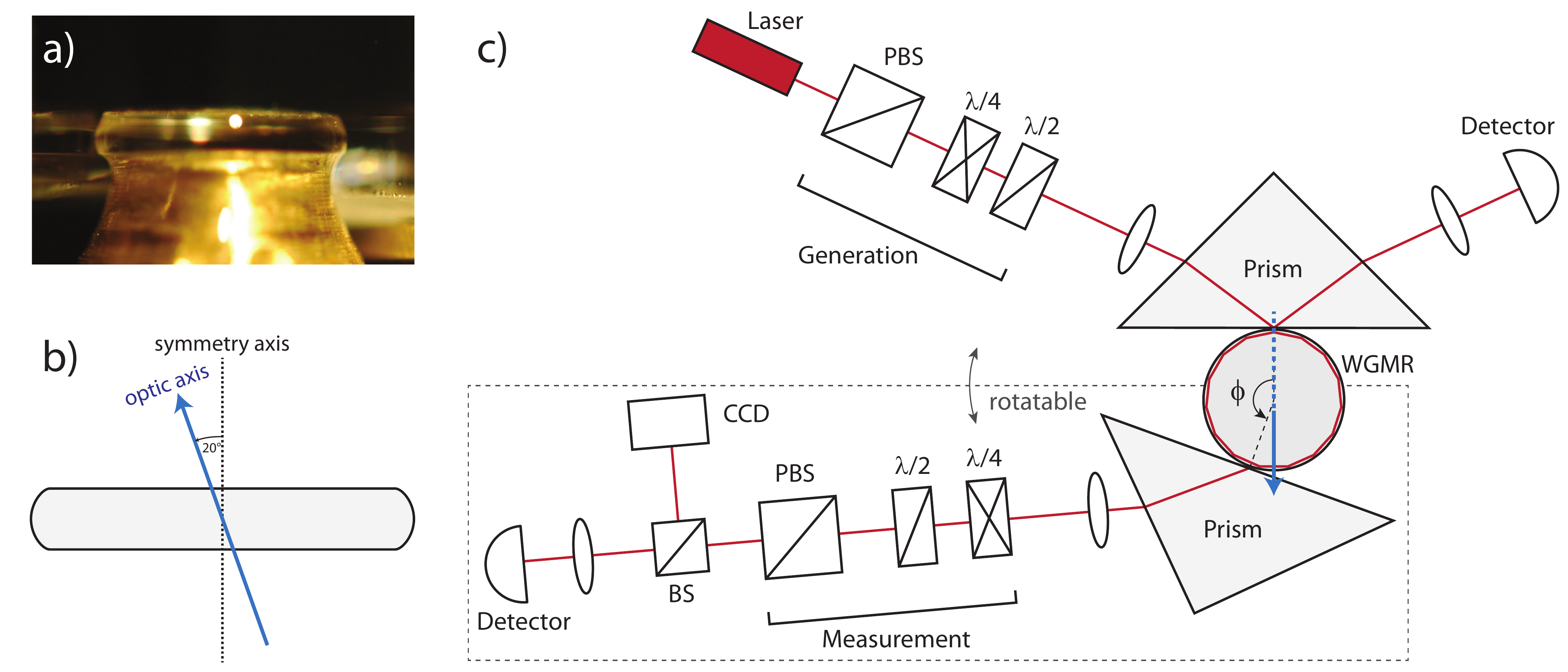}
\caption{a) shows a picture of the \mgf resonator mounted on a brass rod and b) illustrates the schematical cross section through the resonator with the optic axis tilted by $20^\circ$. c) Schematic of the setup: the incoupling port provides full control over the polarization state of the input beam and the cavity spectrum is measured in reflection. The outcoupling prism and the whole Stokes measurement line on the transmission side can be rotated around the symmetry axis of the resonator by an angle $\phi$ over a range of $100^\circ$ from $125^\circ - 225^\circ$. An infrared CCD camera can be optionally mounted to observe the farfield of the outcoupled modes.}
\label{fig:Setup}
\end{figure}

We prepared an angle cut resonator on our in-house single point diamond turning machine. For this purpose a rod was drilled out of a \mgf crystal ($n_o = 1.371, n_e = 1.382$ at $1550 \unit{nm}$). The axis of the rod was tilted by $20^\circ$ against the optic axis of the crystal. We cut the rod perpendicular to its axis into slices of approximately $750\mu$m thickness. The outer surface of one of the resulting \mgf disks was shaped by single diamond turning (measured radius $R = 2.17 \unit{mm}$) such that it supports whispering gallery modes and provides optimized coupling to an external input/output prism \cite{strekalov_efficient_2009}. Due to the tilt of the resonator with respect to the optic axis, the mechanical properties of the crystal surface are not homogeneous along the circumference. Therefore special care needs to be taken during the machining process of such crystals. In case of the comparably soft crystal \mgf we managed to fabricate such a WGM resonator, see Fig.~\ref{fig:Setup}(a, b). In a final step the disk was carefully polished to obtain an optical surface quality to suppress scattering out of the near-surface guided whispering gallery modes.

The obtained whispering gallery resonator was cleaned and mounted between two SF11 prism couplers ($n = 1.743$ at $1550 \unit{nm}$), see Fig.~\ref{fig:Setup}(c). The distance between the couplers and the resonator is piezo controlled (Attocube system) with a precision down to a few nanometers. A commercial tunable laser around $1550 \unit{nm}$ (Toptica DL Pro design) is coupled to whispering gallery modes with the standard prism coupling method. The input polarization is fully controllable due to a polarizing beamsplitter followed by a half- and a quarter-waveplate which enables us to cover any input polarization state on the surface of the Poincar\'e Sphere.

At the incoupling port we can already learn something about the properties of the excited modes by determining the input polarization which gives maximum coupling contrast. Due to the small birefringence of the resonator the coupling does not affect the polarization of the incoming light much, thus the incoupling polarization is expected to be close to the polarization of the excited mode.
As we expect the asymmetry of the system to change the polarization properties as a function of the position of the resonators rim we probe excited modes at various outcoupling positions. For this purpose the outcoupling prism can be rotated around the resonator to extract the light over an angular range of $100^\circ$ (see Fig.~\ref{fig:Setup}(c)). The outcoupled light passes a quarter-waveplate, a half-waveplate, as well as a polarizing beamsplitter to allow for a full Stokes measurement. Finally the beam is split with a non-polarizing beamsplitter, where the intensity is measured at the transmission port with a standard photodiode. A CCD infrared camera is placed at the reflection port to trace the farfield pattern of the transmitted resonator modes and thus enables investigation of possible spatial dependence of the output polarization.

\section{Experiment}

\begin{figure}[t!]
\centering\includegraphics[width=0.95\linewidth]{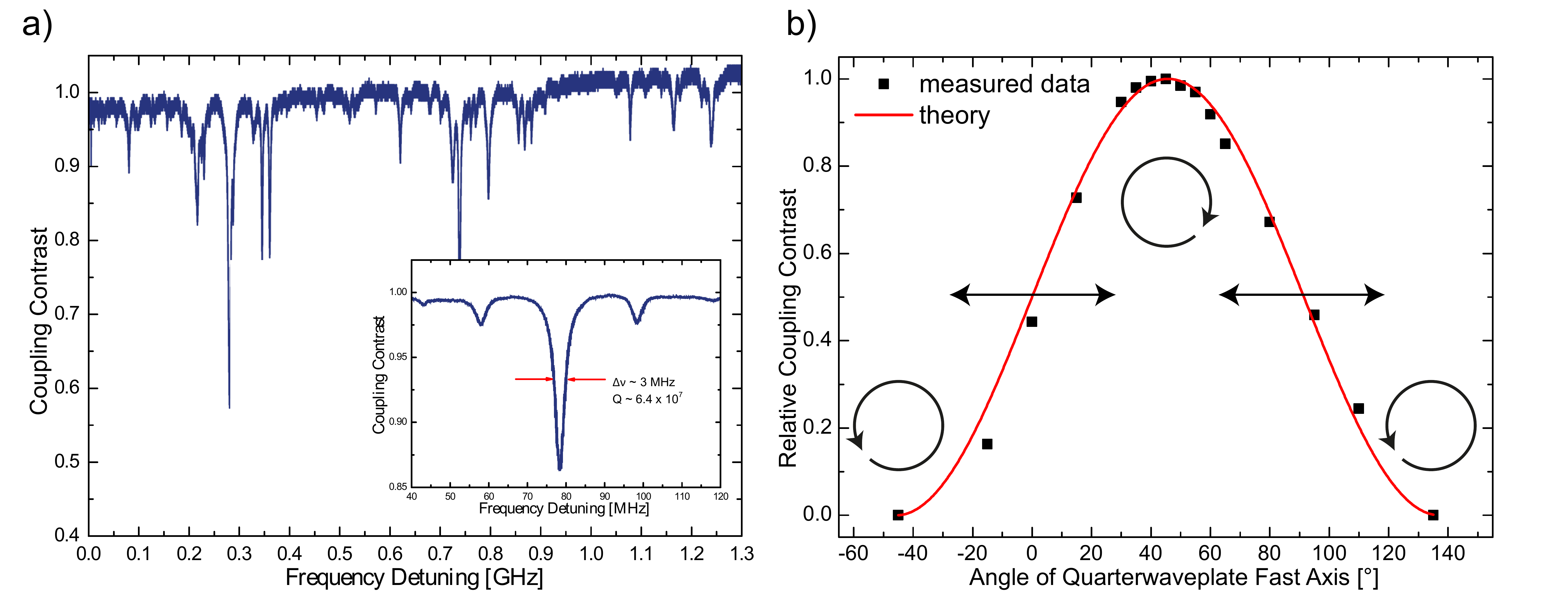}
\caption{a) An overview spectrum from a reflection measurement is shown: the resonator exhibits a dense spectrum with different coupling contrast and slightly varying linewidths. The inset shows a single $Q$-factor measurement with sidebands for frequency calibration. b) shows the measurement of the coupling behavior of a mode which exhibits neither ordinary nor extraordinary properties. The relative coupling efficiency is plotted against the angle of the fast axis of the incoupling quarter-waveplate ranging from $-45^\circ$ (left handed circular polarized) over $0^\circ$ (linearly horizontal polarized) to $45^\circ$ (right handed circular polarized). The comparison with theory (straight line) indicates that the light within the resonator at the coupling position is nearly right handed circularly polarized.}
\label{fig:SpectrumandCircularModeCoupling}
\end{figure}

The observed overview spectra are comparable to those of a common z-cut resonator: they show a dense spectrum of modes of different order and different coupling efficiencies. An overview spectrum is shown in Fig.~\ref{fig:SpectrumandCircularModeCoupling}(a). All those resonances show reasonably narrow linewidth resulting in quality factors of up to $Q\approx6\times10^7$. Though this is a reasonable $Q$ it is much less than the material limit for \mgf WGM resonators \cite{grudinin_ultra_2006}. We were able to reach $Q$ up to $10^9$ out of the same single crystal in a diamond-turned z-cut resonator of similar geometry, thus we have not reached the material absorption limit in our angle-cut resonator yet. As described above, this may be caused by the coupling between the two eigenpolarizations at each reflection of the light opening loss channels.
\begin{figure}[t]
\centering\includegraphics[width=0.9\linewidth]{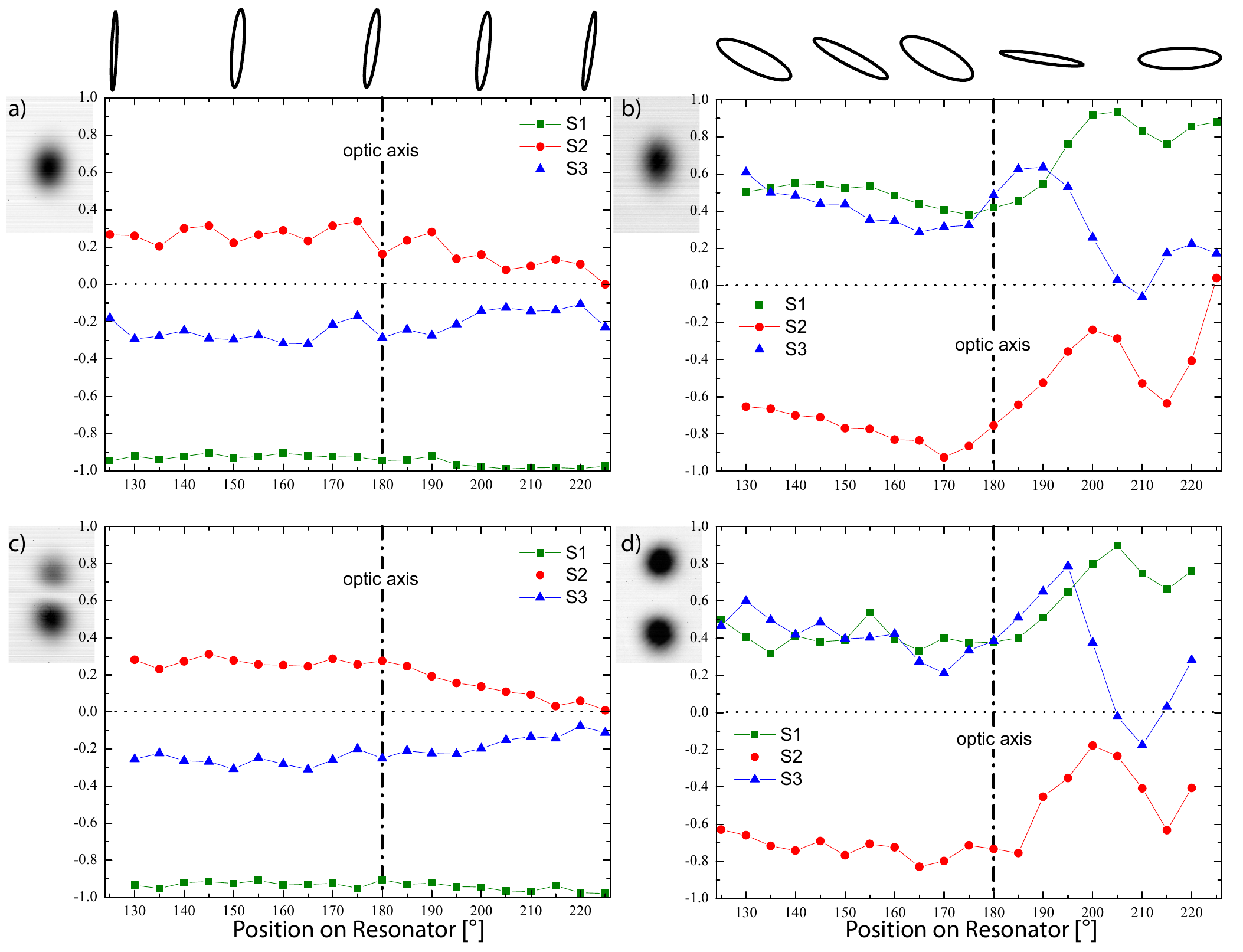}
\caption{The Stokes parameters of the transmitted light in the farfield are plotted against the outcoupling position (see also Fig.~\ref{fig:Setup}(c)). The pictures show for both, extraordinary (a) and (c) and ordinary (b) and (d) polarization, a fundamental and a higher order mode respectively (see camera pictures). One can see that the $l-m$ mode number does not influence the polarization behavior in a qualitative way. The handicity of the polarization ellipses of the two families is different which is reflected by the different signs of $S3$. For illustration Jones polarization ellipses are plotted at five outcoupling points for the fundamental modes.}
\label{fig:StokesFarfield}
\end{figure}

It turns out that the modes can be roughly separated into three categories concerning their polarization behavior. Two types of modes couple best with linear input polarizations which point approximately in the direction of the optic axis or perpendicular to it respectively. Therefore we will call those modes \textit{extraordinary} and \textit{ordinary}. Interestingly it was shown for an angle cut BBO resonator \cite{lin_high-q_2012} that there exists only an ordinary polarized mode. It was argued qualitatively that the extraordinary mode would decay due to the polarization coupling, while the ordinary one survives because of inhibited reflection. As we use a positive crystal in contrast to the negative crystal BBO, this effect should suppress the ordinary polarization direction in our case. Apparently this is not the case for our \mgf resonator, probably because of the by far smaller birefringence compared to BBO.

The third family of modes shows best coupling at various arbitrary elliptical input polarization states. Fig.~\ref{fig:SpectrumandCircularModeCoupling}(b) shows for example the relative coupling efficiencies of a certain mode for different levels of ellipticity: the fast axis of the incoupling quarter-waveplate is adjusted at different angles $\theta$ against the horizontal polarization direction passing the plate. The generated light after the quarter-waveplate is described in circular basis as $\mathbf{E} = E_0/\sqrt{2} \left[( \cos\theta + \sin\theta) \hat{\sigma}_+ + (\cos\theta - \sin\theta) \hat{\sigma}_- \right]$ where $\hat{\sigma}_+$ and $\hat{\sigma}_-$ denote the basis vectors of right and left handed circular polarized light respectively. Thus the intensity coupling contrast to a right handed circular mode as a function of the rotation angle of the quarter-waveplate can be expressed as $I_c / I \propto 0.5 + \cos\theta \sin\theta$. Fig.~\ref{fig:SpectrumandCircularModeCoupling}(b) shows the measured contrast of a mode being in good agreement with this formula and is therefore nearly circulary polarized at the incoupling port. The $Q$ factor of this family remains still above $10^7$, but coupling efficiencies do not exceed 10\%.

In transmission the farfield profile of several modes was investigated and shown to behave mostly in a manner expected for a common z-cut resonator. As shown in Fig.~\ref{fig:StokesFarfield} they produce either a single gaussian or two separated gaussian distributions in the farfield which correspond to either fundamental or higher order $l-m \neq 0$ modes. In spherical geometries the eigensolutions of the wave equation separate and can therefore be addressed by the three modal numbers $l,m,q$, where $l-m+1$ gives the number of intensity maxima along polar direction while $m$ corresponds to the number of wavelengths around the equator and $q$ to the number of intensity maxima along radial direction \cite{oraevsky_whispering-gallery_2002}. In case of anisotropic z-cut resonators the wave equation does not separate anymore. Still perturbative solutions exists for very small deviation and show that field distributions are quite similar to a sphere \cite{ornigotti_theory_2011}. Yet, there is no theory for angle-cut resonators available, but our results indicate that a number of modes of the resonator can still be addressed by the notation taken from the spherical resonator, at least in the case of the tightly bound WGMs. It is also worth to mention that the farfield beam profiles are spatially homogeneous in terms of polarization.

For further investigation we focus on the ordinary and extraordinary modes as they are most interesting for the broadband phase-matching concept described briefly above. 
We measured the free spectral ranges (FSR) of all feasible modes with the sideband method described in \cite{li_sideband_2012}. Assuming the statement that fundamental modes ($q = 1$) have smallest FSR is still true for angle cut resonators we choose the modes showing smallest FSR for ordinary and extraordinary modes respectively. Both pairs contain a mode with fundamental farfield pattern and one with a higher order $l-m \neq 0$ pattern.

\begin{figure}
\centering\includegraphics[height=5cm]{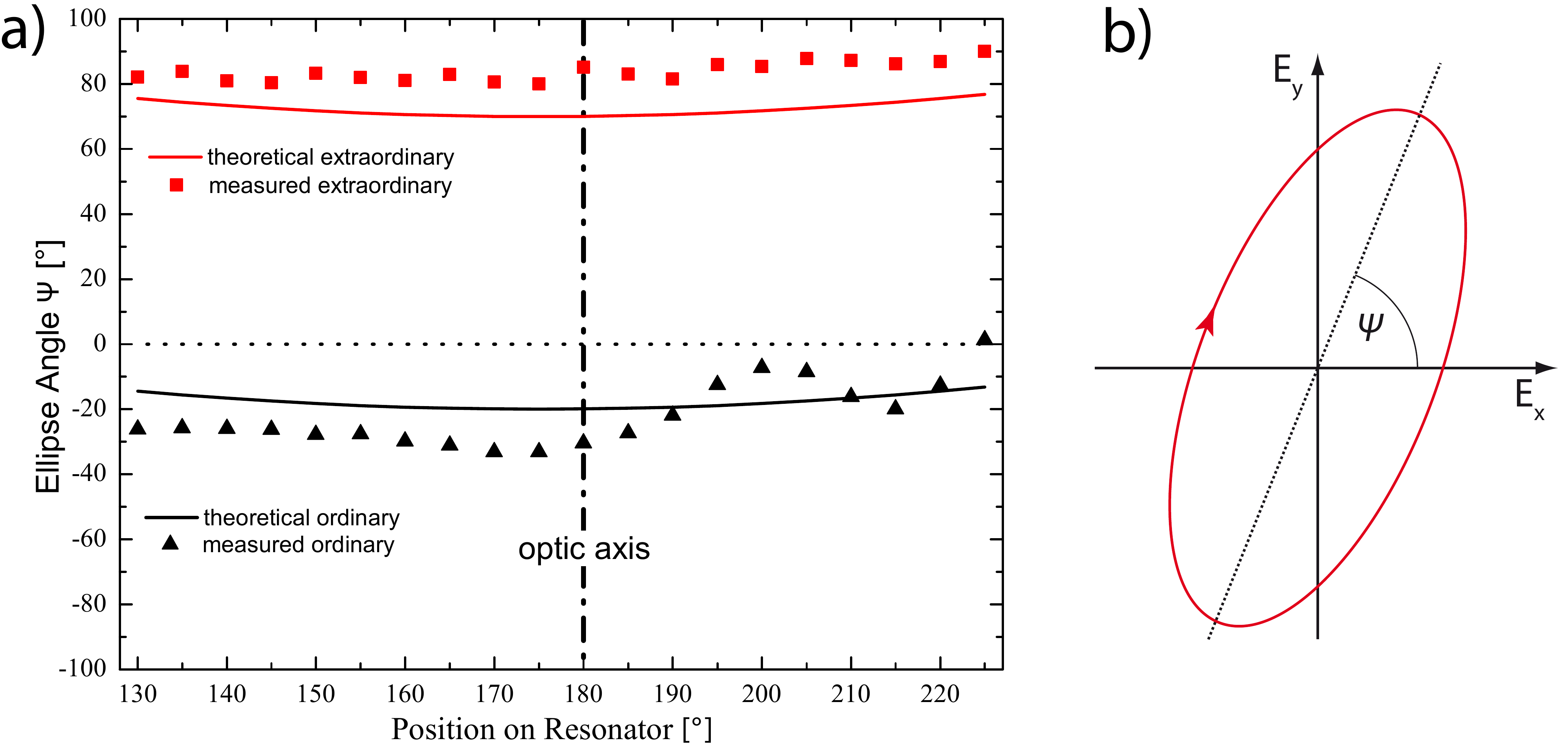}
\caption{a) Angles of the Jones polarization ellipses derived from the Stokes vector set of the fundamental ordinary (black) and extraordinary (red) modes. The straight lines show position dependent angles of actual extraordinary (red) and ordinary (black) linear eigenpolarizations within the resonators bulk material. Those angles vary due to alternation of orientation of k-vector and optic axis while the light travels around the rim.}
\label{fig:AnglesFarfield}
\end{figure}

The results of the measurements covering about $100^\circ$ of the resonators rim are shown in Fig.~\ref{fig:StokesFarfield} in terms of Stokes parameters of the outcoupled light plotted against the angular position. Compared to the z-cut geometry there are two new, unique features of the angle cut resonator: first, the polarization of the outcoupled light depends strongly on the outcoupling position, second, the modes are in general not linearly polarized which is reflected by the non-vanishing $S3$ parameter. Remarkably $S3$ has opposite signs for ordinary and extraordinary modes for the greatest part of the measured range.
Ordinary (Fig.~\ref{fig:StokesFarfield}(b, d)) and extraordinary (Fig.~\ref{fig:StokesFarfield}(a, c)) modes of different spatial profile show, qualitatively, the same polarization behavior respectively. Therefore it makes sense to address them as two different modal families in terms of polarization similar to TE and TM modes in a z-cut resonator.
While the extraordinary modes show a close to linear, weakly varying polarization the extraordinary modes vary more strongly in terms of polarization direction and ellipticity.

Obviously the polarization does not exactly follow the eigenpolarizations of the material which is already clear because of the non-vanishing $S3$ parameter. To obtain a measure for the difference we compare the direction of the major axis of the measured polarization ellipses to the actual eigenpolarizations.
In Fig.~\ref{fig:AnglesFarfield}(a) we show the angles of the major axis of Jones ellipses derived from Stokes parameters against the horizontal direction (see Fig.~\ref{fig:AnglesFarfield}(b) for definition) and plot them against the outcoupling position. The corresponding straight lines represent the angles of calculated linear ordinary and extraordinary eigenpolarizations of the resonator at each position. The directions of the ellipses axis and the eigenpolarizations are  similar but do not coincide. The differences in direction and the ellipticity in particular comes most likely from the coupling of the two eigenpolarizations via the reflection at the birefringent boundary of the resonator.

\section{Conclusion}
Slightly tilted birefringent WGM resonators show a breath of novel features never seen in standard z-cut resonators.  We have shown that high quality resonances still exist in such a geometry and that the polarization states of the resonances are in general no longer linear and vary along the equator of the WGM resonator. Nevertheless, different classes of modes, characterized by their shared polarization properties, can be identified which confirms that there are resonances that experience an varying index of refraction along the equator of the resonator. This offers novel approaches for phase-matching in non-linear optical frequency conversion to extend the possible conversion range in birefringent crystalline resonators. Additionally the handicity of the elliptical part of the observed modes, especially those which show strong S3 values, may open new opportunities for WGM resonator related sensing applications.

\section*{Acknowledgement}
 We acknowledge fruitful discussions with D.V.\ Strekalov and M.C.\ Collodo.

\bibliography{nonZ}
\bibliographystyle{plainnat}

\end{document}